\documentclass[a4paper,UKenglish,cleveref, autoref,nolipics, thm-restate]{lipics-v2021}
\usepackage{amsfonts}

\newcommand{\realrange}[2]{\left[#1, #2\right]}

\newcommand{\unitrange}[2]{\realrange{0}{1}}

\newcommand{\llabel}[1]{\label{\labelprefix:#1}}
\newcommand{\labelprefix}{} %

\newcommand{\discussionsize}{\small}

\newcommand{\frage}[1]{}

\newenvironment{code}{\noindent%
\begin{tabbing}%
\hspace{2em}\=\hspace{2em}\=\hspace{2em}\=\hspace{2em}\=\hspace{2em}\=%
\hspace{2em}\=\hspace{2em}\=\hspace{2em}\=\hspace{2em}\=\hspace{2em}\=%
\kill}{\end{tabbing}}

\newcommand{\labelcommand}{}
\renewcommand{\captiontext}{}
\newsavebox{\codeparam}
\newcounter{lineNumber}
\newenvironment{disscodepos}[3]{%
\renewcommand{\labelcommand}{#2}%
\renewcommand{\captiontext}{#3}%
\sbox{\codeparam}{\parbox{\textwidth}{#3}}%
\begin{figure}[#1]\begin{center}\begin{code}\setcounter{lineNumber}{1}}{%
\end{code}\end{center}\caption{\llabel{\labelcommand}\captiontext}\end{figure}}

{\end{disscodepos}}

\newdimen\endofsize\endofsize=0.5em
\def\endofbeweis{~\quad\hglue\hsize minus\hsize
                 \hbox{\vrule height \endofsize width
\endofsize}\par}

\usepackage{url}
\usepackage{pdflscape}

\newcommand{\neighbors}{\ensuremath{\mathrm{\Gamma}}}%
\usepackage{todonotes}
\usepackage{multicol}
\usepackage{wrapfig}
\usepackage{numprint}
\usepackage{relsize}
\usepackage{graphicx}
\usepackage{wrapfig}
\usepackage{enumitem}

\nolinenumbers

\let\acmcomment\comment
\let\comment\undefined

\usepackage{changes}

\let\comment\acmcomment

\definechangesauthor[name="Ilya Safro", color=red]{is}

\usepackage[left] {lineno}
\definecolor {infocolor} {rgb} {0.6,0.6,0.6}

\newcommand{\ie}{i.e.}
\newcommand{\etal}{et~al.}
\newcommand{\eg}{e.g.}

\npdecimalsign{.} %

\newcommand{\scotchtxt}{\textsc{Scotch}}


\newcommand{\mytitle}{Open Problems in (Hyper)Graph Decomposition}

\authorrunning{Ajwani \etal}
\author{Deepak Ajwani}{University College Dublin, IE}{}{}{}
\author{Rob H. Bisseling}{Utrecht University, NL}{}{}{}
\author{Katrin Casel}{Humboldt University Berlin, DE}{}{}{}
\author{Ümit V. Çatalyürek}{Georgia Institute of Technology -- Atlanta, US, Amazon Web Services, US}{}{}{}
\author{Cédric Chevalier}{CEA, DAM, DIF -- Arpajon, FR}{}{}{}
\author{Florian Chudigiewitsch}{Universität zu Lübeck, DE}{}{}{}
\author{Marcelo Fonseca Faraj}{Universität Heidelberg, DE}{}{}{}
\author{Michael Fellows}{University of Bergen, NO}{}{}{}
\author{Lars Gottesbüren}{Karlsruher Institut für Technologie, DE}{}{}{}
\author{Tobias Heuer}{Karlsruher Institut für Technologie, DE}{}{}{}
\author{George Karypis}{University of Minnesota – Minneapolis, US}{}{}{}
\author{Kamer Kaya}{Sabanci University – Istanbul, TR}{}{}{}
\author{Jakub Lacki}{Google -- New York, US}{}{}{}
\author{Johannes Langguth}{Simula Research Laboratory - Oslo, NO, University of Bergen, NO}{}{}{}
\author{Xiaoye Sherry Li}{Lawrence Berkeley National Laboratory, US}{}{}{}
\author{Ruben Mayer}{Universität Bayreuth, DE}{}{}{}
\author{Johannes Meintrup}{Technische Hochschule Mittelhessen -- Gießen, DE}{}{}{}
\author{Yosuke Mizutani}{University of Utah – Salt Lake City, US}{}{}{}
\author{François Pellegrini}{University of Bordeaux, FR}{}{}{}
\author{Fabrizio Petrini}{Intel Labs -- Menlo Park, US}{}{}{}
\author{Frances Rosamond}{University of Bergen, NO}{}{}{}
\author{Ilya Safro}{University of Delaware -- Newark, US}{}{}{}
\author{Sebastian Schlag}{Sunnyvale, US}{}{}{}
\author{Christian Schulz\footnote{Corresponding Author}}{Universität Heidelberg, DE}{}{}{}
\author{Roohani Sharma}{MPI für Informatik -- Saarbrücken, DE}{}{}{}
\author{Darren Strash\footnote{Corresponding Author}}{Hamilton College -- Clinton, US}{}{}{}
\author{Blair D. Sullivan}{University of Utah -- Salt Lake City, US}{}{}{}
\author{Bora Uçar}{CNRS and LIP ENS de Lyon, FR}{}{}{}
\author{Albert-Jan Yzelman}{Huawei Technologies -- Zürich, CH}{}{}{}
\Copyright{Christian Schulz}
\begin{document}
\title{\mytitle}

\maketitle
 
\begin{abstract}
Large networks are useful in a wide range of applications. Sometimes problem instances are composed of billions of entities. Decomposing and analyzing these structures helps us gain new insights about our surroundings. Even if the final application concerns a different problem (such as traversal, finding paths, trees, and flows), decomposing large graphs is often an important subproblem for complexity reduction or parallelization. 

This report is a summary of discussions that happened at Dagstuhl seminar 23331 on ``Recent Trends in Graph Decomposition'' and presents currently open problems and future directions in the \hbox{area of (hyper)graph decomposition.}

\end{abstract}
\vfill \pagebreak
\tableofcontents
\vfill \pagebreak
\section{Introduction}
(Hyper)graphs provide a versatile framework to depict relationships among various entities. This modeling approach spans across multiple domains, including but not limited to, social media platforms, traffic management, neural networks, and large-scale simulations. However, as data grows exponentially, there's an ever-increasing need for scalable graph-processing techniques. A critical step in many of these techniques is (hyper)graph partitioning, which divides a (hyper)graph into $k$ blocks of nearly identical size while minimizing the inter-block edge count. Emphasizing balance, this partitioning often incorporates a \emph{balancing constraint} ensuring the weight of each block doesn't exceed a certain threshold. In the last four decades, there has been a tremendous amount of research in the area. See for example the book by Bichot and Siarry~\cite{GPOverviewBook},
the survey by Schloegel \etal~\cite{SchloegelKarypisKumar03graph} or  Kim \etal~\cite{KimHKM11} as well as last generic surveys on the topic by
Bulu{\c{c}}~\etal~\cite{SPPGPOverviewPaper} and more recently {\c{C}}ataly{\"{u}}rek \etal~\cite{DBLP:journals/csur/CatalyurekDFGHMSSSSW23}. However, a wide range of challenges remain in the area.
Thus we report  currently open problems and future directions in the area of (hyper)graph decomposition that have been presented during Dagstuhl seminar 23331 on ``Recent Trends in Graph Decomposition''.

\section{Preliminaries}
A \textit{weighted undirected hypergraph} $H=(V,E,c,\omega)$ is defined as a set of $n$ vertices $V$ and a
set of $m$ hyperedges/nets $E$ with vertex weights $c:V \rightarrow \mathbb{R}_{>0}$ and net
weights $\omega:E \rightarrow \mathbb{R}_{>0}$, where each net $e$ is a subset of the vertex set $V$ (i.e., $e \subseteq V$).
The vertices of a net are called \emph{pins}.
We extend $c$ and $\omega$ to sets in the natural way, i.e., $c(U) :=\sum_{v\in U} c(v)$ and $\omega(F) :=\sum_{e \in F} \omega(e)$.
A vertex $v$ is \textit{incident} to a net $e$ if $v \in e$. $\mathrm{I}(v)$ denotes the set of all incident nets of $v$.
The set $\neighbors(v) := \{ u~|~\exists e \in E : \{v,u\} \subseteq e\}$ denotes the neighbors of $v$.
The \textit{degree} of a vertex~$v$ is $d(v) := |\mathrm{I}(v)|$.
 We assume hyperedges to be sets rather than multisets, i.e., a vertex can only be contained in a hyperedge \emph{once}.
Nets of size one are called \emph{single-vertex} nets.
Given a subset $V' \subset V$, the \emph{subhypergraph} $H_{V'}$ is defined as $H_{V'}:=(V', \{e \cap V'~|~e \in E : e \cap V' \neq \emptyset \})$.

A \textit{weighted undirected graph} $G=(V,E,c,\omega)$ is defined as a set of $n$ vertices $V$ and a
set of $m$ and edges $E$ with vertex weights $c:V \rightarrow \mathbb{R}_{>0}$ and edge
weights $\omega:E \rightarrow \mathbb{R}_{>0}$.
In contrast to hypergraphs, the size of the edges is restricted to two. Let $G=(V,E,c,\omega)$ be a weighted (directed) graph. We use \emph{hyperedges/nets} when referring to hypergraphs and \emph{edges} when referring to graphs.
However, we use the same notation to refer to vertex weights $c$,
edge weights $\omega$, vertex degrees $d(v)$, and the set of neighbors $\neighbors$.
In an undirected graph, an edge $(u,v) \in E$ implies an edge $(v,u) \in E$ and $\omega(u,v) = \omega(v,u)$.

\section{Balanced (Hyper)graph Decomposition and Variations}
\subsection{Balanced Hypergraph Partitioning}

\textit{Multilevel Scheme.}
Although traditional coarsening algorithms work particularly well for mesh graphs, their extension to hypergraphs has revealed a lack of understanding and can easily destroy the structure of the hypergraph. It is important to invest in better coarsening techniques tailored specifically to hypergraphs to preserve their structural properties. An interesting avenue in that direction could be incorporating embeddings during coarsening. By leveraging embeddings, coarsening algorithms could potentially achieve better representations of hypergraph structures, leading to improved partitioning outcomes.

While spectral techniques were popular in the pre-multilevel era, pure spectral partitioning was not deemed competitive afterwards -- mainly due to high running times. With the increased performance of today's machines and GPUs, it might be worthwhile to revisit these approaches as multilevel refinement techniques.
Recently, unconstrained refinement (ignoring balance constraint while performing node moves)  with subsequent rebalancing has shown promising results. However, the design space of these types of algorithms is far from being explored.

\textit{Methodology.}
Currently, we are lacking evaluations of real-world applications and workflows that use partitioning for load balancing and communication volume minimization. Therefore, the impact of quality gains in partitioning in terms of running time improvements for the applications are somewhat unclear.
Moreover, in both graph and hypergraph partitioning, we still don't have a uniformly accepted balance constraint definition that works well in the case of vertex weighted (hyper)graphs. There exist definitions that enforce a lower bound on the block weights, add the weight of the heaviest vertex to the balance definition, or simply require that each block must be non-empty. Given that finding a balanced partition is an NP-hard problem even without optimizing an objective function, we should investigate in a balance definition that guarantees the existence of a feasible solution without providing too much leeway in the maximum allowed block weights.

\textit{High-Quality Distributed-Memory Partitioning.}
In recent years, several publications demonstrated that shared-memory partitioning algorithms can achieve the same solution quality as their sequential counterparts. However, the same quality gap still exists between sequential and distributed-memory solvers. 

\textit{Bottleneck Objective Functions.}
For parallel computations, we assign the nodes of a (hyper)graph evenly to processors in a computing cluster. This should balance the computational load across the cluster. However, this does not bound the communication between processors, which can also become a sequential bottleneck if some PEs have to communicate significantly more than others. Therefore, we should investigate in techniques for optimizing bottleneck objective functions.

\textit{The One Partitioning Tool Idea.} 
The graph- and hypergraph partitioning problems come in many different flavors: weighted vs. unweighted (hyper)graphs, directed vs. undirected hypergraphs,  different objective functions, single vs. multi-objective, single vs. multi-constraint, partitioning with fixed vertices, partitioning with variable block weights, etc. Can we join forces and build (upon) a single open-source multilevel framework that is easily extensible to foster the research and development of new partitioning heuristics such that we can have a single tool that actually is able to solve all of these problems?

\subsection{(Hyper)DAG Scheduling}
\label{scheduling}

Let a HyperDAG (or, alternatively, a DAH -- directed acyclic hypergraph) represent a computation and be given by a set of vertices and directed hyperedges, i.e.,
$\mathcal{H}=(\mathcal{V},\mathcal{N})$.
Here, $\mathcal{V}=\mathcal{S}\cup\mathcal{T}\cup\mathcal{O}$
while every directed hyperedge $n\in\mathcal{N}$ consists of a source and an arbitrary number of destination vertices;
i.e., $n\in \mathcal{V}\times \mathcal{P}(\mathcal{V})$, where $\mathcal{P}(\mathcal{V})$ is the power set of $\mathcal{V}$.
The vertices $\mathcal{S}$ are the input (source) data of the computation, the outputs are in $\mathcal{O}$, while intermediate computations are captured by computing \emph{tasks} in $\mathcal{T}$.

Scheduling the computation on a parallel system with $p$ processing units
requires assigning each vertex $v\in V$ a time step $t_v$ and a location $s_v\in\{0,1,\ldots,p-1\}$ that define when and where to execute the intermediate computation in the case of $v\in\mathcal{T}$, or when and where an input (or output) should be available in the case of $v\in\mathcal{S}$ (or $v\in\mathcal{O}$).

Let us initially consider a machine model that costs communication and computation, though does not consider weights for simplicity of presentation -- i.e., each
         data element $v\in\mathcal{S}\cup\mathcal{O}$ uniformly costs some unit storage;
         $v\in\mathcal{T}$ generates intermediate data that costs the same unit storage; and
         $v\in\mathcal{T}$ costs some unit time to compute.

Approaching the scheduling problem from a hypergraph partitioning point of view generates a mutually disjoint $\mathcal{V}_0,\ldots,\mathcal{V}_{p-1}$ partition of $\mathcal{V}$ under some allowed load imbalance $\epsilon$, and minimizes the traditional $\lambda-1$-metric $\sum_{n_i\in\mathcal{N}}\left(\lambda_i-1\right)$; i.e., minimizes the communication volume of data units between parts of the partition\footnote{Here, $\lambda_i$ is defined as the \emph{connectivity} of the $i$th hyperedge, i.e., the number of parts of the partition the vertices in that hyperedge span.}. However, even a perfectly balanced and optimal partitioning may lead to a division of the HyperDAG across the $p$ compute units that exposes no parallelism whatsoever. One solution is to divide the HyperDAG into $s$ layers $\mathcal{L}_0,\ldots,\mathcal{L}_{s-1}$, where vertices $v\in\mathcal{L}_{\leq i}$ are predecessors of those in $\mathcal{L}_{>i}$, and then to partition each layer separately. Recent results show, amongst other results, that the resulting HyperDAG partitioning problem is NP-hard, and also that no polynomial-time approximation algorithm exists~\cite{papp22a}. An additional problem is determining an appropriate $s$, which is a hard problem on its own.

\textit{Optimal Scheduling.}
Similar in motivation to the sparse matrix partitioning problem in Section~\ref{partitioning} in this paper, one challenge is to find optimal schedules for real-world HyperDAGs.
A collection of problems may be found in  open HyperDAG\_DB repository\footnote{\url{https://github.com/Algebraic-Programming/HyperDAG_DB/}},
which welcomes additional problem submissions.
Determining optimal schedules
enables efficient execution of oft-repeated computations, such as those in training neural networks;
enables gauging the effectiveness of current heuristics for online scheduling, such as those used within run-time systems like OpenMP or Cilk; and
enables inspiring better on-line heuristics by looking at optimal examples.
This direction implies finding better ILP formulations and improved pruning strategies for use with optimal scheduling algorithms.
Pruning strategies may furthermore rely on data-driven methods,
see e.g., Juho et al.~\cite{DBLP:journals/heuristics/Lauri0GA23},
trained using entries of the HyperDAG database that have been solved to optimality.

\textit{Models and Hardness.}
The hardness results previously presented depend on assumptions on the underlying machine and cost models.
Indeed, other choices may reveal differing hardness results--
for example, the same recent work shows that removing the layer-wise constraint in favor of a makespan constraint on the HyperDAG partitioning
results in an optimization problem where \emph{evaluating} whether said constraint has been violated
is an NP-hard problem in itself~\cite{papp22a}.
A fundamental challenge thus is to identify what machine and cost model choices
1) result in significantly harder optimization problems,
2) affect the search space underlying the optimization problem and how, and
3) have optimal schedules that relate to one another and how.

Example modeling choices include %
whether time step assignment takes place at the unit vertex granularity or in bulk (e.g., assigning multiple tasks to a single layer);
whether data between vertices are moved individually or in bulk;
whether communication in a time step charges constant latency, a cost proportional to a size, or both;
whether communication size corresponds to volume or $h$-relations\footnote{the maximum of incoming and outgoing messages to or from any partition at a given time step.};
whether communication may overlap with computation; or
whether communication throughput and latency (when costed) are uniform across the $p$ processing units, or instead hierarchical or even topology-dependent.
More detailed initial considerations on such modeling options appear in a pre-print~\cite{papp23a}.

To make each of the three above challenges more concrete, we briefly follow with known examples:
1+2) electing a machine model where vertex-to-time-step assignment happens in bulk and layer-wise, leads to fewer variables in an ILP formulation and thus to a reduced search space; yet, paradoxically, also has stronger known hardness results compared to non-layered hypergraph partitioning~\cite{papp22a};
3) there is at most a factor two difference between optimal BSP solutions\footnote{with BSP, compute tasks and communication are considered in bulk, communication charges both latency and size, communication size is given by $h$-relations, and latency as well as throughput parameters are uniform~\cite{valiant90}.} with overlapping communications versus those without.

\subsection{(Hyper)graph Algorithms and ALP}

Algebraic programming, or ALP for short, enables writing programs with explicit algebraic information passed into the programming framework.
Examples of such algebraic information are binary operators and their properties such as associativity, commutativity, etc.,
as well as richer algebraic structures such as monoids and semirings.
A semiring embodies the rules under which linear algebra takes place,
but allows its generalization to any pair of additive and multiplicative operations under which those rules hold;
for example, while the plus-times semiring enables standard numerical linear algebra,
the min-plus semiring enables shortest-paths computations.
The following two observations are core to GraphBLAS:
        a) most graph algorithms can be expressed in (generalized) linear algebra; and 
        b) our deep understanding of optimizing sparse linear algebra (thus) applies to graph computations.

The recent nonblocking mode of ALP/GraphBLAS performs fusion of linear algebraic primitives under any algebraic structure, at run-time.
It achieves up to $16.1\times$ and $12.2\times$ speedup over the similar state-of-the-art frameworks of Suite\-Sparse:GraphBLAS and Eigen
on ten matrices for the PageRank algorithm,
with similar results for a Conjugate Gradient (CG) solver and sparse deep neural network inference~\cite{mastoras22,mastoras22a}.
Other recent work introduces support for dense linear algebra, matrix structures (e.g., triangular), and views (e.g., permutations or outer products)%
~\cite{spampinato23a}.
It furthermore enables automatic distributed-memory execution of sequential ALP code~\cite{yzelman20,scolari23a}.

\textit{ALP Accelerating Graph Algorithms.}
With the new extensions, both the applicability and performance of the ALP framework has increased,
and should enable the acceleration of graph algorithms that previously only had sequential or otherwise un-optimized representations.
To aid porting efforts,
ALP not only supports the auto-parallelization of linear algebraic formulations of graph problems,
but also that of vertex-centric ones~\cite{yzelman22}.

One challenge is to find graph algorithms that, despite recent advances, remain hard to express using generalized linear algebra,
or graph algorithms that are expressible yet do not achieve high performance.
Examples include $k$-core decomposition and $p$-spectral clustering,
the former done successfully~\cite{li21} and the latter still partially relying on non-ALP code~\cite{pasadakis23a}.

\textit{Graph Algorithms Accelerating ALP.}
Techniques exist to accelerate sparse matrix computations using hypergraph partitioning,
either on distributed-memory~\cite{catalyurek01,vastenhouw05,pelt14},
shared-memory~\cite{yzelman09,yzelman11a},
or both simultaneously~\cite{yzelman14a,yzelman14b}.
However, based on the computation, either the hypergraph representation of a sparse matrix must be adapted,
the minimization objective modified,
or both;
see, e.g., Ballard et al.~\cite{ballard16} who consider sparse matrix--matrix (SpMSpM) multiplication rather than sparse matrix--vector (SpMV) multiplication,
as most preceding cited works.
Thus for ALP as a programming model,
the challenge lies in how these models and optimization techniques may be combined --
preferably transparently to the programmer --
to optimize the arbitrary sequences of computations and inputs that ALP encounters.

For example, while we may readily reuse known techniques to optimize any program consisting of SpMV multiplications with the same input matrix and some vector operations,
the framework must be smart to select a different model and optimization objective when it concerns SpMSpM multiplication instead.
Furthermore,
relying on such existing work requires ALP to translate between different partitionings whenever differing operations or differing input matrices are encountered.

Ideally, however,
the framework co-optimizes across multiple primitives and inputs that it encounters.
The following avenues seem possible:
    1) dynamically building fine-grained (hyper)DAG representations,
           followed by partitioning, (Hyper)DAG partitioning~\cite{10.1007/978-3-030-43229-4_19,doi:10.1137/1.9781611976472.1}, or scheduling (see also the related problem~\ref{scheduling});
    2)     employing a coarse-grained parameterized representation of the computation and employing analytic choices, or
    3)     some mixture of the preceding avenues.
Difficulties with the first solution likely relate to the scale of the resulting optimization problem.
For the second, while work in nonblocking ALP/GraphBLAS execution shows that such approaches may be effective~\cite{mastoras22a},
it is unclear how they may extend to arbitrary primitives and inputs.
A successful solution thus may well lie with the third option,
and require mixed fine- and coarse-grained representations combined with both combinatorial and analytic techniques.

\subsection{Sparse Matrix Partitioning}
\label{partitioning}
Given an $m\times n$ matrix $A$ with $N$ nonzeros, the sparse matrix partitioning problem seeks a partition of $A$ into $p$ disjoint parts $A = \cup_{i=0}^{p-1}A_i$ such that the number of nonzeros in part $A_i$ satisfies $|A_i|\leq (1+\varepsilon)\left\lceil\frac{N}{p}\right\rceil$ for $0\leq i < p$, where $\varepsilon\geq 0$ is a given load-imbalance parameter. Important questions for the area of sparse matrix partitioning as well as graph and hypergraph partitioning are:
\textit{    How good is heuristic bipartitioning compared to exact bipartitioning?
    How good is recursive bipartitioning into $k$ parts compared to direct $k$-way partitioning?}

To answer these questions, we can solve a set of small- and medium-size problem instances to optimality using an exact algorithm either based on the branch-and-bound (BB) approach, or on an integer linear programming (ILP) approach. To answer the first question, a set of 839 matrices has been bipartitioned by the programs MondriaanOpt~\cite{pelt15} and MatrixPartioner~\cite{knigge20}. To answer the second question, an exact bipartitioner has been employed within a recursive bipartitioning program for $k=4$ and it has been compared with an exact direct 4-way partitioner using the program General Matrix Partitioner (GMP)~\cite{jenneskens22} and the commercial ILP solver CPLEX. 

We would like to scale up these initial results, to reach larger problems and be able to answer the main questions with more confidence. Since for $k=2$ the bipartitioner MP works best, we ask whether its algorithm and implementation can be further improved. Parallelization should also help to enlarge our database of solved problems. One might interpret this database as a training set for learning (by either machines or humans) about properties of optimal solutions.

For $k > 2$, we surprisingly found that a basic formulation as an ILP solved by a commercial ILP solver was far superior to the BB solver GMP, and this poses the question what we can learn from the ILP solvers. Furthermore, can we use them for certain types of sparse matrix/graph partitioning?
Finally, can we improve the basic formulation of the ILP to solve even larger problems?

\subsection{Scalable Distributed Memory Partitioning}
Scalability of high quality parallel (hyper)graph partitioning remains an active area of research. In particular, achieving good scalability and quality on large distributed memory machines is still a challenge, but even on shared-memory machines, scalability to a large number of threads seems difficult. Even more difficult is aligning the inherent complexity and irregularity of state-of-the-art algorithms with the restrictions of GPUs or SIMD instructions. Another conundrum is that, for good memory access locality during partitioning, (hyper)graphs need to already be partitioned reasonably well. Hierarchies of supercomputers have to be taken into account during partitioning. This can be done by using multi-recursive approaches taking the system hierarchy into account or by adapting the deep multilevel partitioning approach sketched above to the distributed memory case. When arriving at a compute-node level, additional techniques are necessary to employ the full capabilities of a parallel supercomputer. For example, many of those machines have GPUs on a node level. Recently, researchers started to develop partitioning algorithms that run on GPUs and, while of independent interest, partitioning algorithms developed for this type of hardware can help in that regard. Hence, future parallel algorithms have to compute partitions on and for heterogeneous machines. On the other hand, algorithms should be energy-efficient and performance per watt has to be considered. Lastly, future hardware platforms have to be taken into consideration when developing such algorithms. One way to achieve this will be to use performance portable programming ecosystems like the Kokkos~library~\cite{kokkos}.

\subsection{Balanced Edge Partitioning for Distributed Graph Processing}
The generally accepted formulation of the edge partitioning problem imposes a load balancing constraint on the number of \emph{edges} per partition (cf., e.g., ~\cite{10.1145/3448016.3457300},~\cite{9835611}): $\forall p_i \in P : |p_i| \leq \alpha * \frac{|E|}{k}$ for a given $\alpha \geq 1, \alpha \in \mathcal{R}$, where $|p_i|$ denotes the number of edges in partition $p_i$.  
However, balancing only the number of edges does not always lead to good load balancing in distributed graph processing, as shown in~\cite{10184652}. In some cases, it is better to balance the number of vertex \emph{replicas} -- vertex copies produced whenever incident edges are placed in different partitions.  
The open problem is to achieve an edge partitioning that is balanced both in the number of edges and vertices while minimizing the vertex replication factor. First thoughts in that direction may lead to the formulation of a multi-constraint partitioning problem.

\subsection{Provably Effective Graph/Hypergraph Coarsening}

Today almost all of the state-of-the-art graph and hypergraph partitioning tools utilize multi-level approaches that are comprised of three phases: coarsening, initial partitioning and uncoarsening/refinement. Even though numerous different coarsening techniques have been proposed and many are shown to be effective in multi-level partitioning, we still do not have a provably effective and efficient coarsening technique for graph or hypergraph partitioning problems. The situation is more dire for directed graph and hypergraph partitioning for acyclic partitioning. Keeping acyclicity during coarsening is a desirable property, yet, it is computationally expensive to ensure and maintain acyclicity, with flexible coarsening techniques. 

Today we also do not have a well-defined objective for coarsening. In other words, we do not have well-agreed upon desirable properties of the coarsened graph. Overall we want to solve a partitioning problem, but in multi-level partitioning the overall success of the algorithm is a complex function of the three phases of the multi-level approach. We have many counterexample results showing that the best initial partitioning solution does not always yield the best result. Hence, it is even more difficult to define a goal for coarsening. 

In undirected graph and hypergraph partitioning, many successful tools use randomized heavy edge matching/clustering techniques, where vertices are visited in random order, and they are matched with their unmatched neighbor with the heaviest connection. This randomization helps to ``maintain'' the graph's structure. 
Hence, one potential direction for successful coarsening techniques is randomized algorithms (see Section~\ref{sec:randomized_algs}).

\subsection{Randomized Algorithms for Graph Sparsification and/or Coarsening}
\label{sec:randomized_algs}

Randomized algorithms on networks often involve sampling nodes, edges, or subnetworks~\cite{ganguly2021many,ganguly2021non,10.1145/2743021,spielman2011graph}. 
These sampling techniques are used as subroutines (1) for the solution of fundamental problems on networks, such as connected components, the assignment problem, breadth-first search on long-diameter graphs, or global minimum cut, and (2) for graph learning problems trained with variants of mini-batch stochastic gradient. These problems are widely encountered in the US DOE applications, e.g., the use of the assignment problem in optimal transport (transforming probability distributions) for cosmology, the use of connected components in genomics problems. Sketching and sparsification, which are other randomization techniques used for networks, can be used to find approximate solutions for higher-level problems where the network problem is a subroutine. 
The computational science applications include domain-decomposition solvers, iterative solvers, preconditioning for sparse systems using approximate factorization. Sampling and sketching can also be used as a coarsening technique in multilevel graph partitioning. Unfortunately, the impressive advances in the theory of randomized algorithms for networks has not been translated into practical demonstrations. There are ample research opportunities to bridge this gap between theory and practice, and hence, to produce high-quality software running on modern HPC hardware with demonstrations on application codes.

\subsection{Balanced Streaming Partitioning}
In streaming edge partitioning, edges are presented one at a time in a stream and must be assigned to a partition irrevocably at the moment they are encountered. Degree-based hashing has been proven effective for streaming edge partitioning, however, its potential benefits in the context of streaming \emph{vertex} partitioning remain to be explored. An open problem is to investigate the benefits and challenges associated with using degree-based hashing techniques specifically for streaming vertex partitioning.

Another notable limitation in the current literature on streaming partitioning algorithms is the predominant focus on common ordering strategies for the input (hyper)graph, such as random ordering, breadth-first search, and depth-first search orders. While these ordering strategies provide valuable insights into the performance of partitioning algorithms, they may not fully capture the challenges posed by real-world scenarios. Therefore, it is an open problem to experimentally test streaming partitioning algorithms under adversarial node and edge orderings, particularly in the context of buffered streaming algorithms, where locality has a large impact on the quality of the result.

An open problem in the field of streaming process mapping is to address the problem when the underlying topology cannot be faithfully represented as a hierarchy, but only as a graph or hypergraph. Existing streaming algorithms either ignore the topology, i.e., solve the graph partitioning problem, or optimize directly for hierarchical topologies. However, many real-world scenarios involve complex interconnections that form graph-based topologies.

\subsection{Exact Solvers for Large $k$ (Hyper)graph Decomposition}
\label{sec:scalableexact}

Solving the graph bipartitioning problem to optimality using branch-and-bound algorithms has recently been shown to be highly effective if the optimum solution value is very small. For example, Delling~\etal~\cite{DBLP:journals/mp/DellingFGRW15} can solve instances with millions of vertices to optimality in a reasonable amount of time. It is still open whether such solvers also exist for balanced (hyper)graph partitioning problems where $k$ is significantly larger than $2$. The difficulty here comes from the fact that larger values of $k$ introduce a large amount of symmetry in the problem, i.e.,~if you have some partition of the graph, then any permutation of the block ids is also a partition of the graph that has the same balance and objective. Another problem of current solvers is how to handle dense instances or, more precisely, instances in which the objective function is large.

\subsection{Partitioning Stencils and Analyzing The Performance of Partitioning Tools}
\label{sec:stencils}
We investigate the scalability of the graph- and hypergraph-based sparse matrix partitioning methods in terms of being able to obtain high quality solutions in large problem instances. 
The quality measure that we are interested is the connectivity$-1$ metric, which usually measures the total volume
of communication when vertices represent data items/computations and the hyperedges represent the
dependencies.
Ideally, theoretical investigations help explain the scalability or success of the methods. 
However, the current algorithms in use are too sophisticated to lend themselves to such an approach. 
We are thus looking for sound experimental methodology. 

\begin{figure}
    \centering
    \begin{subfigure}[b]{0.3\textwidth}      \includegraphics[width=\textwidth]{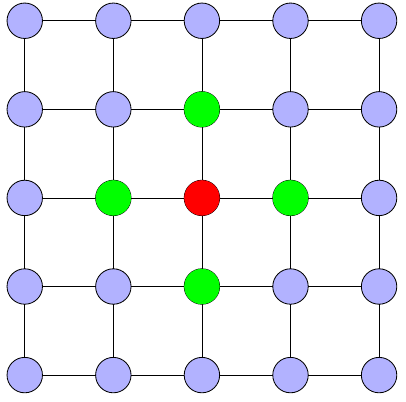}
    \caption{A 2D mesh.}
    \label{fig:2Dmesh}  
    \end{subfigure}
\hspace*{0.1\textwidth}
\begin{subfigure}[b]
    {0.3\textwidth}      \includegraphics[width=\textwidth]{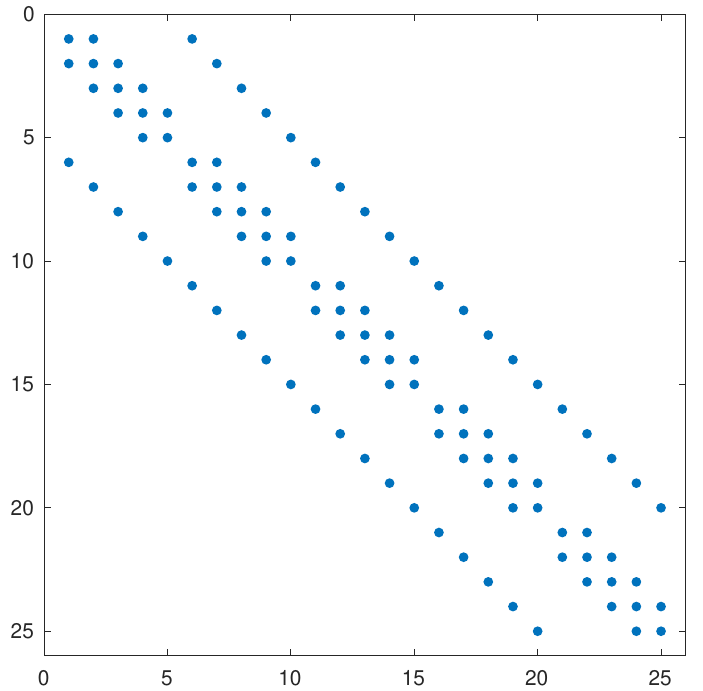}
    \caption{The matrix of the mesh.}
    \label{fig:2Dmeshmatrix}  
\end{subfigure}
\caption{The 2D mesh resulting from the discretization of a square domain with the five-point stencil, using 5 points in each dimension, and the corresponding matrix after a row-by-row ordering of the mesh points. \label{fig:meshandmatrix}}
\end{figure}

One approach is to take a subclass of problems, develop special partitioners, and compare the hypergraph partitioners with those. 
We take five-point stencil computations in two-dimensions (2D) for which we have a special linear-time partitioner~\cite{gruc:14,ucca:10} and see how good the current partitioning methods are on these cases.
We compare the performance of hypergraph partitioners on these. 
A rectangular 2D domain is discretized with the five point stencil and a mesh of size $X\times Y$ is obtained whose points are placed at integer locations. 
Two points $(x_1 , y_1)$ and 
$(x_2 , y_2)$ of the resulting mesh are neighbors iff 
$|x_1-x_2|+|y_1-y_2| = 1$. 
A sample mesh resulting from the discretization of a square domain with five points in each dimension is shown in Figure~\ref{fig:2Dmesh}. 
The figure also shows the connections of a point.
After an ordering of the mesh points, one can obtain an $X^2\times Y^2$ matrix. 
The matrix obtained from the shown mesh after a row-major ordering is shown in Figure~\ref{fig:2Dmeshmatrix}.

A partitioning of the points of the mesh $\mathcal{M}$ corresponds to a row-wise or a column-wise partitioning of the associated matrix $\mathbf{A}_{\mathcal{M}}$. 
Without loss of generality let us focus on the row-wise partitionings of $\mathbf{A}_{\mathcal{M}}$.
The standard column-net hypergraph $\mathcal{H}_{CN}$ model~\cite{caay:99} can be used for this purpose.
Partitioning the vertices of the hypergraph $\mathcal{H}_{CN}$ among $K$ parts will therefore correspond to partitioning the stencil computations among $K$ processors; 
the connectivity$-1$ metric of the cut will measure the total communication volume;
and the balance of part weights in terms of vertices will correspond to balance of the loads of the processors. 
We assume unit vertex weights here, the effect of having less operations on the border points of the mesh will be ignored for simplicity (and is negligible).

\begin{figure}
    \centering
    \begin{subfigure}[t]{0.3\textwidth} \includegraphics[width=\textwidth]{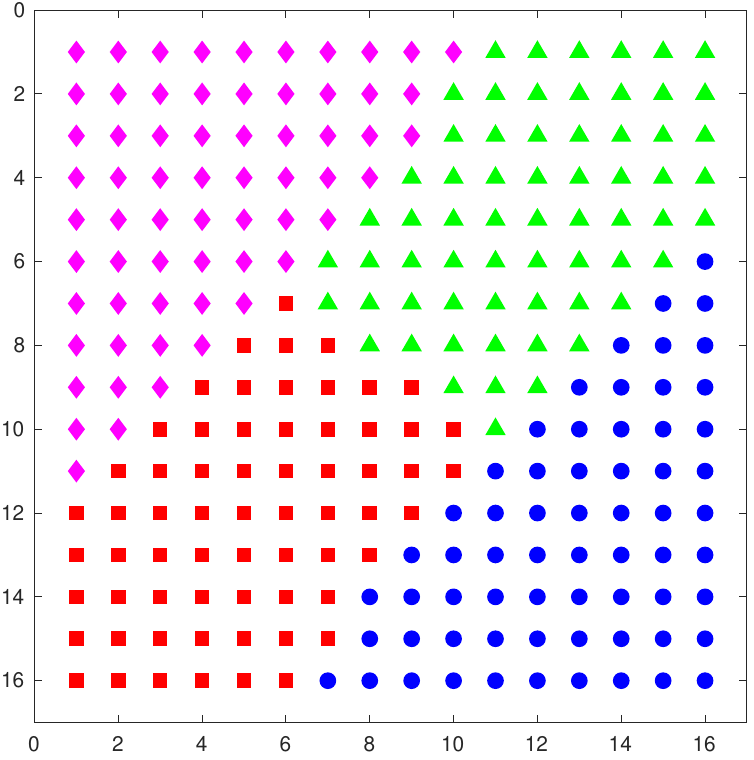}
    \caption{The $K=4$-way partition of the $16\times 16$ obtained by the special routines~\cite{gruc:14}. Connectivity$-1$ is 58.}
    \label{fig:16x16ag}  
    \end{subfigure}
\hspace*{0.01\textwidth}
\begin{subfigure}[t] {0.3\textwidth}    \includegraphics[width=\textwidth]{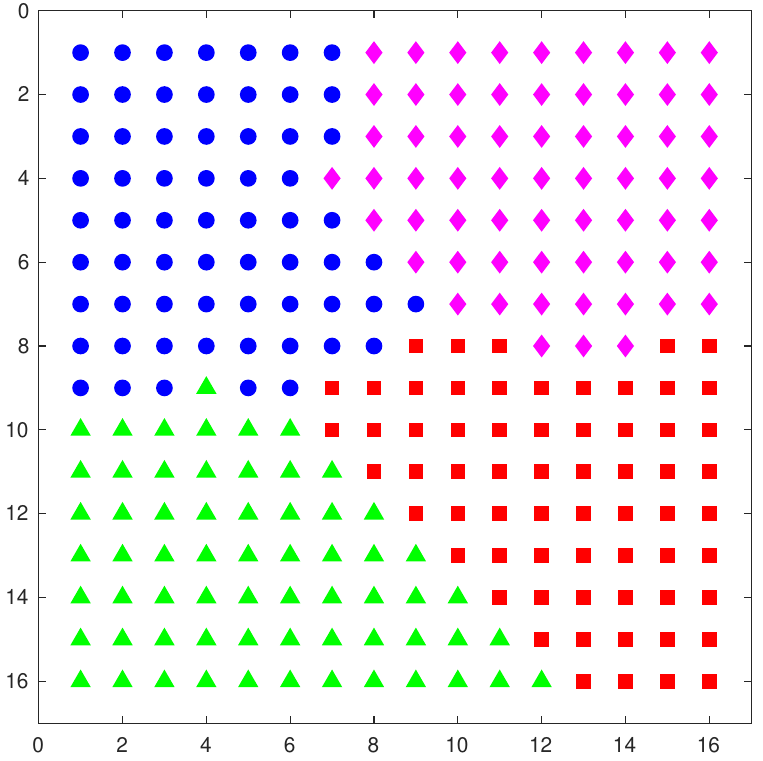}
    \caption{The $K=4$-way partition of the $16\times 16$ mesh obtained by PaToH, suboptimal (connectivity$-1$ is 61).}
    \label{fig:16x16patoh}  
\end{subfigure}
\hspace*{0.01\textwidth}
\begin{subfigure}[t]{0.3\textwidth}    \includegraphics[width=\textwidth]{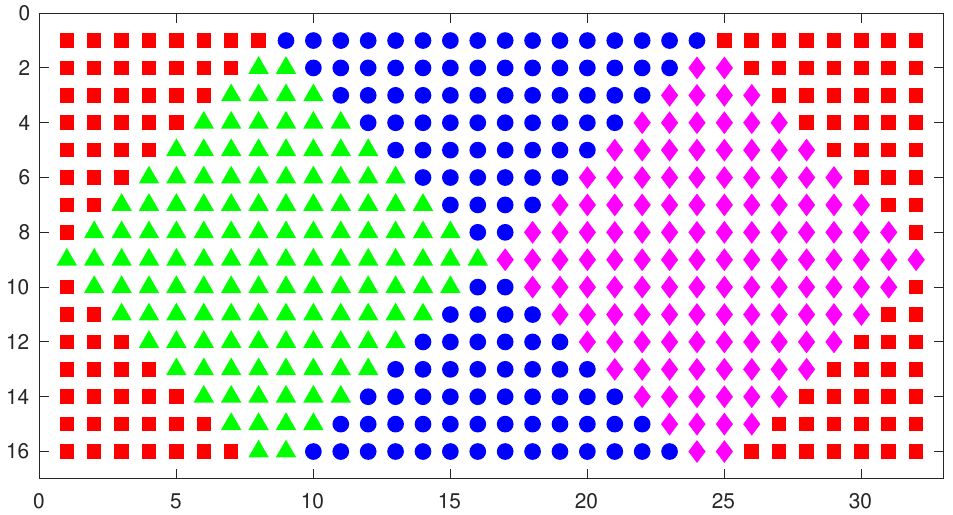}
    \caption{The $K=4$-way partition of the $16\times 32$ mesh by the basic diamonds~\cite[Section 4.8]{biss:04}, which is believed to be asymptotically optimal. Here though the connectivity$-1$ is 126 and PaToH obtains 93.}
    \label{fig:16x32diamonds}  
\end{subfigure}
\caption{Partitioning the $16\times 16$ mesh with different routines; the basic diamonds cannot partition this mesh, but can partition a larger one\label{fig:parts}.}
\end{figure}

With the special partitioning methods by Grandjean and U\c{c}ar, we obtain the total communication volume listed in Table~\ref{tbl:claimCut}.
While the communication volume listed in the table is obtained with the routine itself, 
we note that it is given by the formula 
\begin{equation}\label{eq:isOptCut}
2\times \left(\left\lfloor\frac{n}{\sqrt{2}}\right\rfloor+n\right)+4\;, 
\end{equation}
which is claimed to be optimal for $X>16$ in the table (for $X=16$, the optimal communication volume is claimed to be 57).

This formula requires some conditions on $X$, which we do not give here. Are the connectivity$-1$ values given in Table~\ref{tbl:claimCut} optimal for a perfectly balanced 4-way partitioning of the two mesh of $X\times X$ points discretizing a square domain with a five point stencil? For the $16\times 16$ mesh, Gurobi solver~\cite{gurobi} also finds a cut of value 57 in a few minutes under an additional constraint to assign the four corners to four different parts).  The solution obtained by the Gurobi solver is at the close neighborhood of what is shown in Figure~\ref{fig:16x16ag}; 
by the method of Grandjean and U\c{c}ar is easily updated to mimic Gurobi's result.

Another point that arise from the given 4-way partitioning is that these partitions cannot be obtained in a recursive bisection scheme where each step greedily optimizes the cut hyperedges with perfect balance.
This is so, as the first cut vertically cuts the mesh into two equally sized parts with perfect balance and optimal cut. 
Simon and Teng~\cite{site:97} delve more into this point in the context of graph partitioning.

We note that more results of the sort are given elsewhere~\cite{gruc:14}. The same reference also surveys some results from the literature, including references on discrete isoperimetric problems~\cite{wawa:77}, which can be used to guide algorithms.
Two things are of particular note: in the corners, the optimal parts are triangle-like, as in two corners in Figure~\ref{fig:16x16ag}, and in the interior the optimal parts are diamond-like as in Figure~\ref{fig:16x32diamonds}.

\begin{table}
\begin{center}
\begin{tabular}{r|rrrrr}
$X$ & 16 & 64 & 256 & 1024 & 2048\\\hline
PaToH & 61 & 249 &985&4034&7999\\\hline
Vol &  58 &222 &878 &3500 & 6996 
 \end{tabular}
 \end{center}
\caption{The partitioning of the $X\times X$ mesh of five-point stencils with perfect balance using the method of Grandjean and U\c{c}ar~\cite{gruc:14} obtains the numbers given in row ``Vol'' as connectivity$-1$ metric of the cut. PaToH's results are in the row ``PaToH''--- they do not have perfect balance.
Apart from 58 in the last line, the other numbers are claimed to be optimal.\label{tbl:claimCut}}
\end{table}

Bisseling and McColl~\cite{bico:94}  propose digital diamonds to partition similar meshes with wrap-around connections. 
Digital diamonds are  
$\ell_1$-spheres defined with a center $(c_x, c_y)$ and a radius $\rho$. 
Such a diamond contains all mesh points $(p_x, p_y)$ where with $|p_x-c_y|+|p_y-c_y|\leq \rho$.
Grandjean and U\c{c}ar give formulas for the total communication volume when one uses digital diamonds.
They also specify conditions on mesh and part sizes under which a partition by digital diamonds are possible.
Basic diamonds proposed by Bisseling~\cite[Section 4.8]{biss:04} trim off two borders from the digital diamonds to address partitioning of another set of mesh and part sizes---these conditions as well as the total volume of communication are also specified by Grandjean and U\c{c}ar.
Digital diamonds and basic diamonds are believed to be asymptotically optimal in terms of the total communication volume, 
but obtain disconnected partitions on the borders of the mesh when 
there are no round-around connections---which is not desirable in certain applications.

Another approach is to take general sparse matrices boost the data in a way, reason about it and evaluate the performance of hypergraph partitioners on these. 
For example, suppose we partition a matrix $A$ row-wise into $k$ parts, and obtain a total communication volume of $T_V$ units in SpMxV. Then, if we partition the matrix $B=[A, A]$ row-wise again into $k$ parts, then the first partition should be good for $B$ with $2\times T_V$ communication volume. If our partitioning tool is good, such a performance is expected; if the answers were not $T_V$ vs $2\times T_V$, we could either improve the partition of $A$ or $B$. Similarly, a $k$-way row-wise partition of $C=[A; A]$---this time two copies of $A$ are stacked to have twice as many rows---should have a communication volume of $T_V$ units. 
Some experimental investigation with PaToH~\cite{patoh:manual} using these matrix repetition schemes~\cite{ucca:10} reveals a good behavior.
What else can we say about the behavior of partitioning tools on more general problems?

\subsection{Highly Spread Out Weights in Mesh Partitioning}
\label{sec:mesh-partitioning-weights}

Many distributed numerical simulations rely on mesh partitioning to improve the balance of their computations on every computing unit, thus increasing their efficiency and scalability.  
A mesh is modeled by its dual graph or hypergraph as input to a partitioner: each vertex corresponds to a cell of the mesh, and the vertex weight is the computing cost of this cell. Good quality hexahedral meshes have a reasonably regular topology, mostly looking like a 2D or 3D grid. However, the vertex weight distribution can be highly spread out for various applications like Monte-Carlo particle transport simulations. For this kind of instance, classic multi-level approaches of the existing graph partitioner can have some quality issues, symmetric to the ones observed in Section~\ref{sec:stencils}.
Multi-level graph partitioners focus on topological properties, and here, taking more into account vertex weight distribution should lead to better and faster obtained partitions.

\subsection{Cartesian Mesh Partitioning}
\label{sec:mesh-partitioning-cartesian}

Directly addressing data in memory is crucial in achieving high performance when running on modern architectures, especially on GPU. Grids allow direct access to neighbor cells for mesh computations, making stencil computations like the one presented in Section~\ref{sec:stencils} very efficient.
However, standard partitioning approaches lead to non-rectangular parts, making distributed applications less efficient. Thus, a new problem is partitioning a grid into parts that are all a subgrid or a set of subgrids.
Such a partitioning model will also work to partition block structured meshes that often arise for hexahedral meshes.

\subsection{Problems in Multilevel Graph Partitioning With Star Graphs}

The partitioning community has long focused on instances with a regular structure, e.g., mesh graphs or instances from circuit design. However, it becomes more and more important to find high-quality solutions for instances with an irregular structure, such as those derived from social networks. Surprisingly, we found a subclass of these instances where current state-of-the-art partitioning algorithms compute solutions that are far from optimal. The identified instances -- referred to as \emph{star instances} – are characterized by a core of a few highly-connected nodes (core nodes) with only sparse connections to the remaining nodes (peripheral nodes). 

One example of such an instance is the \textsc{Twitter} graph. Here, we found that partitioning the nodes into low- and high-degree vertices ($\le$ median degree) induces a bipartition that cuts half the edges as any of the existing multilevel partitioning tools. We identified several other social networks where we observed the same behavior. Thus, it becomes increasingly important to develop efficient partitioning techniques that can handle such instances.

From a theoretical perspective, we were already able to present an $(R+1)$-approximation for star instances, where $R$ is the ratio of an approximation algorithm for the min-knapsack problem. This is a remarkable result since there exists no constant factor approximation for the general graph partitioning problem.

\subsection{Graph Partitioning with Ranked Vertices}

For some graph algorithms, there is an implicit rank over the vertices. For instance, 2-hop indexing generates a {\em label cover} that can be used for answering pairwise shortest-distance queries. Classical algorithms, e.g., Pruned Landmark Labeling~({\sc Pll})~\cite{Akiba2013} and its variants, leverage a ranking that has a drastic impact on the number of entries stored at local vertex indexes. In the distributed setting, the amount of entries in the {\em cut}, i.e., the ones replicated and/or communicated among the nodes, depends on this ranking. Especially when the number of nodes is high, this communication can create a bottleneck. For distributed execution, ranking the vertices also changes the loads on each part. In that sense, another problem at hand is given the rank, the amount of data stored at each vertex needs to be estimated well enough so that the part weights incur an acceptable level of imbalance. Hence, the problem is given a graph $G = (V, E)$, {\em what is the best vertex ranking and partitioning pair that yields the best performance in terms of execution time and maximum memory used at each node?}

\subsection{Designing Multilevel Algorithms}
In a variety of fields, computational optimization challenges often arise when modeling large and complex systems, presenting significant hurdles for solving algorithms, even when high-performance computing resources are deployed. These obstacles are frequently due to a multitude of factors such as an extensive number of variables and the complexity in describing each variable or interaction. Problems involving combinatorial and mixed-integer optimization add extra layers of complexity. Specifically, the presence of integer variables frequently results in NP-hard problems, particularly in contexts where nonlinearity and nonconvexity are factors. 

A widely adopted strategy for tackling these challenges involves the use of iterative algorithms. While these algorithms may be grounded in divergent algorithmic paradigms, they often exhibit a similar pattern: rapid improvement during initial iterations followed by a phase of slower progress. In the realm of iterative algorithms, utilizing first-order optimization techniques like gradient descent or methods that rely on limited observable data, such as local search, often leads to a local optimum that is usually suboptimal when compared to the true global optimum. Additionally, the algorithms employed within each iterative cycle are not always exact, further complicating the optimization process. To speed up these algorithms at each iterative step, various strategies including heuristics, parallelization, and different ad hoc techniques are commonly employed, albeit often at the expense of solution quality. Being trapped in local optimum of unacceptable quality is one of the most important issues of such algorithms. 

Multilevel methodologies have been introduced to address the challenges of large-scale optimization, offering a strategy that reduces the chances of being trapped in low-quality local optimum. These techniques are complementary to stochastic and multistart approaches, which also help the algorithm escape local optima. While there's no one-size-fits-all prescription for designing multilevel algorithms, their core philosophy revolves around global considerations while executing local actions based on a hierarchy of increasingly simplified representations of the original complex problem.

In practice, a multilevel algorithm initiates the optimization process by generating a hierarchy of progressively simplified (or coarser) problem representations. Each subsequent coarser level aims to approximate the problem at the current level but with fewer degrees of freedom, facilitating a more efficient solution process. After solving the coarsest problem, its solution is extrapolated back to the more detailed level for further refinement -- a phase termed as ``uncoarsening.'' Employing this multilevel approach frequently results in substantial improvements in both computational efficiency and the quality of solutions. There are many broad impact open questions in designing multilevel algorithms for (hyper)graphs some of which we mention here.\\

\noindent {\it Distance between vertices.}  In order to coarsen the problem, a critical issue is to design a distance (or similarity) function between nodes. The question is simple: how to introduce a similarity function that will effectively find subsets of nodes that share the same solution (e.g., in the context of graph partitioning it is about predicting that nodes will be assigned the same part)? Incorrectly chosen subsets of nodes will mislead coarsening and will make the uncoarsening to work much harder which will result in increased complexity and poor results. In the same time, sophisticated distance functions are not supposed to destroy the overall complexity of the multilevel algorithm. Examples of such advanced solutions are spectral-based~\cite{chen2011algebraic,ron2011relaxation,shaydulin2019relaxation} and low-dimensional representations \cite{sybrandt2020hypergraph}. They work very well on the partitioning, ordering \cite{safro2011multiscale} and clustering multilevel schemes. However, there is also a lot of evidence that these algorithms are not perfect and do not fit all scenarios.\\

\noindent {\it Density of coarse levels.} This remains one of the most crucial issues in multilevel algorithms. In many problems and coarsening schemes the more we coarsen the problem, the more dense graphs are obtained unless we deliberately take actions to sparsify them. On the one hand, such dense representations often may approximate the original problem better. On the other hand, the complexity of refinement at the corresponding levels of uncoarsening becomes prohibitive. For example, in the algebraic multigrid inspired multilevel approach for graph linear ordering this issue was simply patched by reducing the interpolation order \cite{safro2009multilevel} which is a pretty blind solution. It was slightly improved in graph partitioning \cite{safro2015advanced} by using a better node distance function in combination with the small interpolation order but more sophisticated and theory-grounded approaches are required. In a similar 2-dimensional layout problem, the authors switched to more regular coarsening with the geometric multigrid \cite{ron2010fast}. In general, dense graphs are problematic for most existing multilevel algorithms that mostly designed for sparse instances and require special treatment such as other coarsening schemes or special hardware \cite{liu2022partitioning}.\\

\noindent {\it Maximization problems.} A particularly interesting class of problems for which multilevel algorithms have not reached their advanced stage is maximization problems such as max cut, maximum independent set, and maximum dominating set. A traditional coarsening approach quickly generates dense coarse levels and becomes impractical. Recent work on sophisticated node distance functions and sparsification improve the situation \cite{angone2023hybrid} but after a certain number of levels the quality of coarse levels becomes either poor (if sparsified) or intractable (otherwise). Rethinking of the coarsening ideas is required for this class of problems as such approaches as inverting graphs quickly become impractical.\\ 

\noindent {\it When to stop the refinement?} Perhaps there is no multilevel algorithm whose developers have not asked this question. Overall, there is no theory-grounded work related to optimization on (hyper)graphs that answers this question. Apart from the complexity issue, on the first glance it may look trivial that in the ideal refinement, the employed local optimization solvers should be optimal. However, there is a lot of practical evidence that terminating refinement before reaching the best possible local solution is beneficial to the final global suboptimal solution. \\

\noindent {\it Advanced types of multilevel cycles.} In multilevel schemes, the V-cycle coarsening-uncoarsening is the most basic and widely used cycle for this purpose, but several other advanced cycles aim to improve the efficiency and effectiveness of multilevel methods. Most widely used of them are: (1) the W-cycle is a more advanced version of the V-cycle that  provides a more aggressive approach to solving the coarser problems. In a W-cycle, a refinement and full deeper W-cycle is performed at each coarser level before moving back to the finer level. This allows for more thorough refinement at lower levels, often leading to better convergence properties compared to the V-cycle. (2) The F-cycle method creates the hierarchy of coarse representations and starts at the coarsest level and works its way up to the finest grid, solving the problem at each level by applying another full V- or W-cycle. It combines with V-cycles or W-cycles at each level for better optimization of coarse levels. Both F- and W-cycles are particularly effective for problems where an initial coarse approximation is not easy to obtain. Both cycles usually exhibit better then in V-cycle quality which comes at additional cost of complexity. The W-cycles are usually more expensive but do exhibit a good quality~\cite{safro2006graph,safro2006multilevel}. Finding robust criteria on when to recursively apply one or another type of advanced cycle (if at all) is very important in multilevel algorithms as their running time is increased with the advanced cycles. 
\vfill \pagebreak
\section{(Hyper)graph Clustering}
\subsection{Correlation Clustering} 
In the correlation clustering problem the input is a graph with edges labeled with $+$ and $-$ (or simply with $+1$ and $-1$).
$+$ indicates that the endpoints of the edge should be in the same cluster, and $-$ means that the endpoints of the edge should be in different clusters.
The goal of correlation clustering is to find a clustering that respects as many of these requirements as possible.
Of course respecting all of them is in general not possible, and so a commonly studied objective is to minimize the number of disagreements.

There is a big discrepancy between the theory work on correlation clustering and what is done in practical solutions.
For example, while the famous PIVOT algorithm provides $3$-approximation for complete graphs, if the algorithm is run on a sparse graph (i.e., one where $+$ edges induce a sparse graph) the algorithm often gives a solution that is worse than leaving each node in a cluster of size $1$.
Better approximation algorithms are known, but they are not as scalable, as they rely on solving an LP or SDP.
In the case of weighted or not-complete graphs the best known approximation ratio is $O(\log n)$.

Despite all of these theoretical advances, the solutions that are implemented in practice are based on local swaps and a multilevel approach.
In particular, the basic operation that these algorithms make is moving a node to a neighboring cluster, only if this increases the overall objective.
This way, the algorithm essentially treats the objective function as a blackbox and does not leverage all the structural properties of the problem, which are used to give approximation algorithms.

While the practical implementations are quite scalable, there is probably room for improvement, as the number of logical rounds needed to obtain a good solution goes in hundreds.
This in particular makes these algorithms not easy to use in distributed settings.

An interesting open problem is to bridge the gap between theory and practice for correlation clustering with the goal of obtaining better practical implementations.
Specifically, it would be interesting to develop algorithms requiring fewer rounds, which will make them amenable to an efficient distributed implementation.

\subsection{Overlapping Edge-Colored Clustering}

\textsc{Edge-Colored Clustering} is a
categorical clustering framework~\cite{ANGEL201615} whose input is an edge-colored hypergraph and output is an assignment of colors to nodes which minimizes the number of edges where any vertex has a color different from its own (\emph{mistakes}).  We are interested in variants of this problem which allow budgeted overlap. Specifically, the following three notions were defined in \cite{crane2023overlapping}.
     \textsc{LocalECC} allows up to $b$ of color assignments at each node.
     \textsc{GlobalECC} allows one ``free'' color assignment for each node,
    plus $b$ additional assignments
    across all nodes.
     \textsc{RobustECC} allows each node to either receive exactly 1 color, except that at most $b$ nodes are assigned \emph{every} color. Equivalently, at most $b$ nodes are deleted.

Each of these problems generalizes \textsc{ECC}, with equivalence for the first coming at $b = 1$ and for the latter two at $b = 0$. Consequently, they are each NP-hard (Angel et al.~\cite{ANGEL201615}). We (Crane et al.~\cite{crane2023overlapping}) showed that greedy algorithms give an $r$-approximation on the number of edge mistakes, where $r$ is the
maximum hyperedge size. Further, for \textsc{LocalECC}, a $(b+1)$-approximation can be achieved with LP-rounding. 
More generally, we ask about  bicriteria $(\alpha, \beta)$-approximations, where $\alpha$ is the approximation factor
on edge mistakes and $\beta$ is the approximation factor on the budget $b$, and show that all three variants have such approximation algorithms, though the factors are no longer constants for \textsc{GlobalECC}.

\textit{Open Problems:} 
     Are these ideas relevant for any practical applications? Where? What can we assume about the inputs in those settings?
     Are there constant-factor single-criteria approximations for the Global and Robust versions? Does \textsc{GlobalECC} have a constant-constant bicriteria approximation? More generally, bicriteria inapproximability is an interesting and relatively unexplored direction.
      We saw that empirically these approximations performed much better in practice than the guarantees. Is there some sort of structure in real-world instances that we can model to improve our analysis? 
 
\subsection{Dense Graph Partition}
\textsc{Dense Graph Partition}, introduced by Darley et al.~\cite{DBLP:journals/dam/DarlayBM12}, models finding a community structure in a social network. Formally, given an undirected graph $G=(V,E)$, the task is finding a partition $\mathcal{P} = \{P_1, \dots, P_k\}$ of $V$, for some $k \geq 1$, of maximum density.  With $E(P_i)$ denoting the number of edges among vertices in $P_i$,  the density of $\mathcal{P}$ given by $d(\mathcal{P})=\sum_{i=1}^k \frac{|E(P_i)|}{|P_i|}$. 

Note that there is no restriction on the number of communities which yields some difference to the problem of partitioning into cliques. While there exists a partition into exactly $k$ sets of density $(n-k)/2$ if and only if the input graph can be partitioned into $k$ cliques~\cite{DBLP:conf/swat/BazganCC22}, there can be a partition into less than $k$ sets with a density higher than $(n-k)/2$ even if the input cannot be partitioned into $k$ cliques.
Alternatively, \textsc{Dense Graph Partition} can be modeled from a game theoretic perspective. 
Aziz et al.~\cite{DBLP:conf/ijcai/AzizGGMT15} study the \textsc{Max Utilitarian Welfare} problem where the vertices in a graph $G=(V,E)$ are agents, and each agent $x\in V$ validates its coalition $P\subseteq V$ with $x\in P$ by  $\frac{1}{|P|}|\{u\in P\mid \{u,v\}\in E\}|$. Maximizing social welfare for this model is equivalent to \textsc{Dense Graph Partition}.

It is known that maximum matching is a 2-approximation~\cite{DBLP:conf/ijcai/AzizGGMT15}, and there are a few improvements on specific graph classes: polynomial-time solvability on trees~\cite{DBLP:journals/dam/DarlayBM12}, $\frac 43$-approximation on maximum degree 3 graphs, and EPTAS for everywhere dense graphs~\cite{DBLP:conf/swat/BazganCC22}. This in particular gives rise to the questions: Can the 2-approximation be improved, at least on some more non-trivial graph classes?  Does there exist a polynomial-time approximation scheme on general instances?  Is it true that there is always an optimum solution where all parts induce a graph of diameter at most 2, a so-called \emph{2-club clustering}?  What is the complexity on graphs of bounded treewidth?

\subsection{Streaming Graph Clustering} 

\textsc{Streaming Graph Clustering} is commonly defined as follows: given a graph $G=(V,E)$, find a clustering $\mathcal{C}:V \rightarrow \mathcal{N}$ that maximizes a quality score such as modularity, using at most $O(|V|)$ memory. In the one-pass version, $E$ is an ordered list of edges and each edge can be read only once.
A popular heuristic for this problem is SCoDA \cite{hollocou2017linear}.

This matches well with real-world applications where graphs are discovered over time, e.g.~in online social networks, as well as for graphs which are too large to cluster using standard $O(|E|)$ memory algorithms. However, the one-pass version is quite limiting and often results in low clustering quality \cite{langguth2021incremental}.

The \textsc{Incremental Graph Clustering} model \cite{langguth2021incremental} is a \textit{buffered} variant of the one-pass model where the ordered edge list is subdivided into \textit{batches}. Unlike in buffered streaming graph partitioning \cite{faraj2022buffered}, the batches are assumed to be given and not selected by the algorithm. Edges in a batch are read and processed in memory together. The algorithm can use $O(|E|)$ memory, but we require that running time for processing each batch does not depend on $|E|$, only on the size of the batch.

The Neighborhood-to-community link counting (NCLiC) \cite{langguth2021incremental} is a heuristic for this variant.  For modularity clustering, it provides strong modularity retention compared to offline algorithms. It applies the Leiden Algorithm to each new batch and then merges it with the already processed graph.

\textit{Open Problems:} 
The NCLiC algorithm keeps track of the approximate number of neighbors in each cluster. If a vertex changes community it will
update its neighbors with a probability that depends on its degree. Skipping some updates allows maintaining the required running time,
but introduces a reduction in clustering quality. Is there a data structure that allows keeping exact counts of the neighboring clusters without violating the running time constraint?

Another open question is: is it possible to modify NCLiC to use at most $O(|V|)$ memory while retaining most of the modularity retention qualities?

\section{Data Reductions and Learning}

\subsection{Data Reductions for (Hyper)Graph Decomposition}
Most balanced (hyper)graph partitioning formulations are NP-hard: it is believed that no polynomial-time algorithm exists that always finds an optimal solution.  However, many NP-hard problems have been shown to be fixed-parameter tractable (FPT): large inputs can be solved efficiently and provably optimally, as long as some problem parameter is small.  Over the last two decades, significant advances have been made in the design and analysis of fixed-parameter algorithms for a wide variety of graph-theoretic problems. Moreover, in recent years a range of methods from the area have been shown to improve implementations drastically. For example for the maximum (weight) independent set problem \cite{DBLP:conf/gecco/GrossmannL0S23}. Here, data reductions rules transform the input into a smaller one that still contains enough information to be able to recover the optimum solution. For the maximum independent set problem, this enabled highly scalable exact solvers that can solve instances with millions of vertices to optimality. For balanced partitioning this has currently not been carefully investigated. However, here are some very simple data reduction rules. For example, removing a vertex of degree one, then solving the smaller subproblem with same balance constraint and afterwards assigning the vertex to a block with leftover capacity, is a valid data reductions rule. This yields the natural open questions: are there more and highly effective data reduction rules for balanced (hyper)graph partitioning problems? These rules could be helpful in two ways: they could speedup current heuristic solvers, e.g.~multilevel (hyper)graph partitioning algorithms, and they could help to build more scalable exact partitioning algorithms (see Section~\ref{sec:scalableexact}).

After all reduction rules for kernel computations have been applied, the final smaller instance can still be too large to be solved to optimality within a reasonable time bound. This is a serious problem as the overall goal of the algorithms is to solve the given problem instance. 
The idea of lossy kernelization is as follows: when no more reductions can be applied, i.e.~a problem core has been computed, one may shrink the input further while guaranteeing that the optimal solution value changes only slightly. Then a good approximate solution of the reduced input can be lifted to a good approximate solution of the original input. This has recently been done for the vertex cover problem \cite{DBLP:conf/alenex/LavalleeRSP20}. The natural question that arises is can these techniques be applied to balanced (hyper)graph partitioning as well? 

As the (lossy) kernel/core still contains the optimum solution (or some approximation thereof) in some sense, this has a large potential to speed up the (multilevel) heuristic while not sacrificing solution quality. Additionally, running a fast algorithm on the large kernel can help to identify parts of the instances that are likely to be in a good solution. Those parts can then be put into a partial solution and the remaining instance can be reduced recursively. 
 
It could also be possible to use machine learning to learn lossy reductions for a wide-range of problems in this area. For example, one could use learning to predict if two vertices should be clustered together or to decide if an edge is a cut edge or not. The basic idea is then to use a classification model to learn which parts of the input can be pruned, i.e. are unlikely or highly likely in an optimum solution. In the first case, a solution omits this part of the input, in the latter case this part of the input will be included in the solution. For example, \cite{DBLP:journals/heuristics/Lauri0GA23} propose to use machine learning frameworks to automatically learn lossy reductions for the maximum clique enumeration problem and \cite{TayebiRA22} shows that this learning-to-prune framework is effective on a range of other combinatorial optimization problems. The classification model can be a deep neural network in an end-to-end framework or a classifier with significantly fewer parameters such as SVM or random forest if a deeper integration of machine learning and algorithmic techniques is done. The latter will require carefully engineered features based on existing heuristics.

\subsection{Learning for Local Search in Multi-level (Hyper)graph Partitioners}

Machine learning techniques can also be used to learn more efficient refinement steps. Existing refinement steps in multi-level graph partitioning techniques rely on solving a flow problem or iterative moves of Kernighan--Lin or Fiduccia–Mattheyses heuristic. However, solving flow problems can be quite slow (given the number of times it is called). Similarly, the number of possible moves that need to be explored for finding a good step using Kernighan--Lin or Fiduccia–Mattheyses can be quite high. It is worthwhile exploring if learning techniques can be used to predict good regions where the flow algorithm can focus. This can improve the trade-off between the time to solve the flow problem and the gain from it for the refinement part. For the case of the Kernighan--Lin or Fiduccia–Mattheyses heuristic, the interesting question is whether learning techniques such as reinforcement learning can be used to learn a good sequence of moves for these local search heuristics. This has the potential to reduce the search space that needs to be explored to find good local moves. 
For training the learning techniques, the R-MAT graph generator from the Graph500 benchmark can be used.

\section{Embeddings}

\subsection{Distance Estimation for Process Mapping}

Process mapping is a super-problem of graph partitioning, in which vertices of some source graph $S$ have to be assigned (\ie, mapped) to vertices of some target graph $T$, by way of a mapping function $\tau_{S,T}: V(S) \longrightarrow V(T)$, so that an objective function is minimized. In the field of parallel computing, source graphs commonly represent computations to be performed, usually multiple times in sequence, while target graphs represent processing elements and interconnection networks of multi-processor and/or multi-computer hardware architectures. The objective function to minimize is the amount of data to be exchanged across the interconnection network, so as to reduce its congestion, provided that every processing element in $V(T)$ receives roughly the same number of vertices of $S$ (or, more generally, equivalent vertex weights with respect to its compute power), to minimize computation imbalance. In this context, partitioning some graph $S$ into $k$ parts amounts to mapping $S$ onto $K(k)$, the complete graph of order $k$, since in this case all processing elements are at the same distance from all the others.

In the Dual Recursive Bipartitioning (DRB) algorithm~\cite{pellegrini296682} used by the \scotchtxt\ software, computing the mapping of $S$ onto $T$ requires to be able to estimate the shortest-path distance in $T$ between any two vertex subsets of $V(T)$ called the \textit{subdomains} of $V(T)$. These subdomains are not arbitrary, since they result from recursively bipartitioning the graph $T$ into pieces of roughly the same size in a way that minimizes the cut of the interconnection network. Being able to compute the distance between any two subdomains allows the DRB algorithm to estimate the penalty of assigning some vertex $v$ of $S$ to either one of two sibling subdomains of $T$, by estimating the distance between these subdomains and those to which all the neighbor vertices $v$ of $u$ have already been mapped. When the recursive bipartitioning of $T$ is perfectly balanced, the number of subdomains of $T$ is $2 |V(t)| - 1$.

A way to quickly obtain the distance between any two subdomains of some target graph $T$ is to pre-compute a distance matrix between all of them, of a size in $O(|V(T)|^2)$. While this solution works for small target graphs, it is no longer applicable when mapping onto big parts of very big target architectures.
To solve this problem, one has to find a more compact (in terms of data storage) and quick (in terms of retrieval time) method to produce these distance estimates. An important condition on these approximations is that distances should become more accurate as subdomains are smaller and closer to each other in $T$.

In \cite{pellegrini:hal-01253509,pellegrini:hal-01671156}, it has been shown that, for target architectures for which the recursive bipartitioning of subdomains, and the distances between subdomains, can be computed algorithmically, by way of explicit functions (\eg, for regular vendor architectures such as meshes, butterfly graphs, etc.), a bipartition tree, created by way of recursive matching and coarsening of the whole target graph, allows one to represent any subset, even disconnected, of the processing elements of these target architectures. The DRB algorithm can therefore be applied to them.

However, for irregular architectures (\eg, those represented by irregular graphs), the question remains open. It can be expressed in the following form: “How can one get cheaply (both in terms of memory and computation time) approximate distances between any pair of the subgraphs yielded by the recursive bipartition of some irregular graph?”

\subsection{Space-Efficient Planar Graph Embedding}

When one opens up a publication regarding planar graph
bisections, one often reads a sentence akin to:
Without loss of generality, assume that the input graph
is embedded in the plane and maximal planar.
Famous works that makes use of this specific property
is the balanced separator theorem due to Lipton et al.~\cite{LiptonTE80}, which 
states that every planar graph has a balanced vertex
separator of size $O(\sqrt{n})$. Standard recursive bisection
algorithms for
planar graphs are based on this theorem, which are
able to construct the entire recursive bisection in linear time~\cite{KleinMSS13}.
Often it is easy to assume such an embedding,
as it can be computed in linear time using $O(n \log n)$
bits of space, i.e., a linear number of words.
In sub-linear space settings one can compute an embedding
in polynomial time (albeit with an extremely large polynomial degree). Now, the question remains: what can one achieve
when aiming for a linear time algorithm, while using
$o(n \log n)$ bits, or ideally, $O(n)$ bits? The standard linear time algorithms
are quite involved, but on the most basic level many
of them use a simple depth-first tree and compute a constant
number of, but seemingly critical, variables per vertex.
Even when aiming for a much lower goal: 
check if the input graph is planar within $O(n \log n)$
time while using $o(n \log n)$ bits, there is no obvious way to 
tackle this problem. As graphs grow larger and larger,
such questions of space-efficiency become of higher interest.
Especially with the direct application of graph partitioning
algorithms that rely on such embeddings.
\vfill
\pagebreak

\subsection{Finding Moore-Bound-Efficient Diameter-3 Graphs}

In graph theory, given a graph with degree $d$ and diameter $k$, the largest number of vertices in that graph can be determined using the Moore bound.  Recent technological advances in photonics technology have greatly increased the number of links -- or degree $d$ -- of the network routers, improving the scalability of large supercomputers.  While Moore-bound-optimal diameter-2 graphs have recently been engineered to span a few thousand nodes~\cite{DBLP:conf/sc/LakhotiaBMIIHP22}, emerging AI and graph applications are demanding larger configurations. Unfortunately, diameter-3 graphs are still elusive, with Moore’s bound efficiencies of only 15\%.  
The construction of more efficient diameter-3 graphs would directly impact the design of emerging photonics systems for large scale graphs~\cite{DBLP:conf/sc/LakhotiaBMIIHP22,DBLP:journals/corr/abs-2302-07217}, data analysis, and AI applications.

\section{Parameterized Complexity}

\subsection{Parameterized Complexity of Layered Giant Graph Decomposition}

\noindent \textit{Direction 1:}   An important theme -- or challenge -- for theoretical computer science, that has been recognized for decades, is the observation that has been made prominently by Richard Karp and others that we don't really understand very well natural input distributions. \textit{It is remarkable how well sometimes very simple heuristics work in practice for problems that are known to be NP-hard.}  There must be some sort of structure, but what is it?  And if we knew, could we exploit that in designing algorithms? 

A striking example of this was described by Karsten Weihe in an old paper entitled, "On the Differences Between Practical and Applied" which was about Weihe's experience doing quite practical computing for a simple  \textsc{Hitting Set}  application in real-world computing where his project was tasked with computing a minimum number of stations that could service all of the trains of Germany.  

The model is a straightforward bipartite graph, with trains on one side, and stations on the other, and an edge if a train $t$ stops at a station $s$.  There are two simple pre-processing vertex deletion rules: (1) If $N(s)$ is a subset of $N(s')$, then delete $s$. (2) If $N(t)$ is a subset of $N(t')$ then delete $t'$.  Weihe found that these two simple reduction rules cascade back-and–forth on the gigantic real-world train graphs, and one ends up with (using PC terminology) a kernelized instance that consists of disjoint connected components that have size at most around 50, so the problem can be solved optimally by analyzing the connected components separately.

From the standpoint of parameterized complexity theory, we could simply declare the structural parameter of interest to be: $k$ =``the maximum connected component size of the network $G'$ that results when opportunities to apply the two reduction rules have been exhausted''.  This would be perfectly legal in the mathematical framework of PC --- we could call the parameter the Weihe-width of the  \textsc{Hitting Set}  instance and have a pretty good FPT algorithm for computing what we could call a Weihe-width decomposition.  This is legal, but from a traditional parameterized algorithms and complexity perspective, not entirely satisfying.  

At the expense of quadratic blowup one can combinatorially reduce the very important medical- and bio-informatics problem of \textsc{Feature Selection} to \textsc{Hitting Set} in the following natural way.  We now have enormous amounts of information concerning the genes that are being expressed into RNA, and so each patient in our hypothetical hospital has a gene activation profile.  And each patient either does, or does not, have cancer.

We want to know a small subset of the genes to pay attention to so that we can accurately predict the outcome.  On the one side, we have a vertex for each pair of patients that have differing outcomes, and on the other side, we have one vertex for each gene.  It is surprising that Weihe's two reduction rules work quite practicably, in this very different real-world large data context.  What is going on, and how can we generalize? 
\noindent \textit{Open Problem:}
Can a Weihe-width $k$ decomposition of a graph of size $n$ be computed in truly linear FPT time? 

A second open problem begins by reconsidering the most central example of an FPT graph problem,  \textsc{Vertex Cover}, that has inspired in various ways a surprising amount of theoretical work in the parameterized complexity research community. For example, the recent work reported at IJCAI 2020 on the parameterized complexity \textit{paradigm of solution diversity} began with an initial FPT result about the naturally parameterized  \textsc{Diverse Vertex Cover}  problem.

A (parameterized) vertex cover of a graph is a set of $k$ vertices $V'$ of $G = (V, E)$ such that the largest connected component of $G '= G - V'$ is size one! In other words, deleting the vertices of $V'$ kills off all the edges of $G$, yielding, if we want to call it that, a very nice decomposition of $G'$ into clusters of extremely high data-integrity and coherence, as each connected component consists of a single vertex.

It might seem that the  \textsc{Vertex Cover} problem is so simple that it might be irrelevant for giant graphs. But by setting thresholds for declaring edges, it has been used very effectively in stages in very large dataset bioinformatics, e.g., Dehne's CLUSTAL~W package for multiple sequence alignment \cite{DBLP:conf/iccsa/CheethamDPRT03}.

A key point is that the  successful CLUSTAL W algorithm begins by decomposing a sparse graph constructed by making edges between vertices (data objects) that are emphatically NOT similar. The impulse would be to seek cliques of compatible vertices, but here is exploited that the naturally parameterized  \textsc{Clique}  and \textsc{Vertex Cover} problems are parametrically dual, and from that point the CLUSTAL W algorithm proceeds in stages with an initial decomposition based on a vertex cover cutset on a sparse graph based on thresholding the NOT similarity that makes  an edge in the initial graph.

The following problem explores a generalization where the resulting connected components (“clusters”) satisfy other simple integrity requirements. 
It is called \textit{Vertex Decomposition into Small Dominator Clusters}: given a graph $G = (V, E).$ and parameter $(k, d)$, the question is can we delete  $k$ vertices from $G$, obtaining $ G'$ such that every connected component of $ G'$ has domination number at most $d$?
It is interesting to start by asking if this might be FPT for the vector parameter $(k, d)$. But, if we fix $k = 0$, then the problem is W[2] - complete. We can still hope for a parameterized tractability result, where $d$ is allowed to play an XP-role in the exponent of the polynomial and for fixed $d$, with parameter \hbox{$k$ we get FPT.}
\textit{Open Problem:}
Is this FPT?  And if so, can the corresponding decomposition be computed in truly linear FPT time for $d = 1$?

Note that we could define endlessly many interesting and largely unexplored parameterized problems in a similar manner where the decomposition is modeled by connected components formed by essentially a cutset.
And there is also the possibility of interestingly layered decompositions of this kind.  For example in the \textit{Layered Vertex Cover} problem: given a graph $G = (V, E).$ and parameter $(k, k', k'')$ the question is can we delete  $k$ vertices from $G$, obtaining $G'$ such that every connected component $C$ of $G'$  has the property that: $k'$ vertices can be deleted from $C$ resulting in a graph $C''$  such that each connected component of $C''$ has a vertex cover of size at most $k''$?

Or perhaps our particular application intention might be naturally served by deleting $k$ vertices so that the resulting connected components have nice properties governed by a parameter $t$, and these components can be further decomposed into connected components with a different nice property governed by $t'$ and so on. \\

\noindent\textit{Direction 2:}  The theme of fairly simple and elemental decompositions based on vertex- and edge-cutsets is important.\\

\noindent\textit{Direction 3:}  Since the size $n$ of the networks (graphs) targeted in this application area is huge, the attention should be focused on \textit{truly linear-time FPT}, that is, processing that is simply of $O(n)$ cost, regardless of any parameterization $k$ that we might want to consider.  Polynomial time $O(n^c)$ with no exponential costs associated to the parameter $k$, is the best kind of FPT. For very large graphs, we need $c=1$. 
Slightly more generally, we might consider reasonable FPT processing time-costs of the form $O(n^c + f(k))$, where again the exponent of the polynomial part is $c = 1$, which we will call truly linear time FPT. This is an area of PC structural complexity theory little explored.  There is a small amount of relevant recent work by Jianer Chen and coauthors.

\subsection{FPT Approximation of Vertex Bisection}

Edge (resp.~Vertex) Bisection is one the fundamental graph partitioning problems, where given a graph $G$ and an integer $k$, the goal is to find a set of at most $k$ edges (resp.~vertices), say $S$, such that the vertex set of $G \setminus S$ can be partitioned into two almost equal parts $V_1$ and $V_2$, that is $||V_1| - |V_2|| \leq 1$, and there are no edges between a vertex of $V_1$ and $V_2$, that is $E(V_1,V_2) =\emptyset$.

In the regime of parameterized complexity, Edge Bisection admits a fixed-parameter tractable (FPT) algorithm parameterized by the solution size $k$. In particular, it admits an algorithm running in time $2^{O(k \log k)} n^{O(1)}$~\cite{DBLP:journals/talg/CyganKLPPSW21}, where $n$ is the number of vertices in the input graph. 

\vspace{2mm}
\noindent\textit{Open Problem 1:} One can solve Edge Bisection, for fixed $k$, in linear time? Preferably, is there an algorithm solving Edge Bisection in $2^{O(k \log k)} n$-time?

\vspace{2mm}
\noindent\textit{Open Problem 2:} Does Edge Bisection admit an algorithm with running time $2^{O(k)} n^{O(1)}$? Or can one show that there is no algorithm for this problem that runs in time $2^{o(k \log k)} n^{O(1)}$ under reasonable complexity assumptions?
\vspace{2mm}

\noindent In contrast to Edge Bisection, Vertex Bisection is known to be W[1]-hard~\cite{DBLP:journals/tcs/Marx06}, that is it is unlikely that it admits an FPT algorithm parameterized by $k$. 
On the kernelization front, Edge Bisection cannot admit a polynomial kernel under reasonable complexity assumptions~\cite{DBLP:journals/mst/BevernFSS15}.
This leads of interesting questions regarding the fixed-parameter tractability and/or kernelization \emph{with approximations} for these problems.
In particular, the following questions remain intriguing.

\vspace{2mm}
\noindent\textit{Open Problem 3:} Does Edge Bisection admit a polynomial $\alpha$-lossy kernel, for some $\alpha > 1$? Are there lossy reduction rules that help in solving the problem practically?

\vspace{2mm}
\noindent\textit{Open Problem 4:} Does Vertex Bisection admit an FPT-approximation algorithm? That is, in time $f(k,\epsilon) n^{O(1)}$, can one find a set of at most $(1+\epsilon)k$ vertices, say $S$, such that $V(G\setminus S)=V_1 \uplus V_2$, $||V_1| - |V_2|| \leq 1$ and $E(V_1,V_2) =\emptyset$, or report that there is no such set $S$ of size at most $k$? 

\vspace{2mm}
\noindent\textit{Open Problem 5:} In scenarios where Vertex Bisection pops up in practical usage, can we identify some structure on the instances? For example, can we say that the graphs in interesting instances belong to some nice graph class, or ``is close to'' being in a graph class (this could, for example, be formalized using distance to triviality measures), or have some bounded parameter. If such an identification is possible, studying these scenarios theoretically may lead to interesting insights about the problem.

\subsection{FPT in Decomposition}

\textit{The first open question here is if \textsc{Densest $k$-Subgraph} FPT parameterized by modular-width?}
Given a graph $G$ and an integer $k$, the \textsc{Densest $k$-Subgraph} problem asks for a subgraph of $G$ with at most $k$ vertices maximizing the number of edges.
It is known that this problem is FPT by stronger parameters such as neighborhood diversity and twin cover, yet it is W[1]-hard by weaker clique-width.
Modular-width is defined using the standard concept of modular decomposition \cite{modular-width-gajarsky}.
    Any graph can be produced via a sequence of the following operations:
      (O1) Introduce: Create an isolated vertex.
      (O2) Union $G_1 \oplus G_2$: Create the disjoint union of two graphs $G_1$ and $G_2$.
      (O3) Join: Given two graphs $G_1$ and $G_2$, create the complete join $G_3$ of  $G_1$ and $G_2$.
          That is, a graph $G_3$ with vertices $V(G_1) \cup V(G_2)$ and edges $E(G_1) \cup E(G_2) \cup \{(v, w) : v \in G_1, w \in G_2\}$.
      (O4) Substitute: Given a graph $G$ with vertices $v_1,\ldots,v_n$ and given graphs $G_1,\ldots,G_n$, create the \textit{substitution}
          of $G_1,\ldots,G_n$ in $G$. The substitution is a graph $\mathcal{G}$ with vertex set $\bigcup_{1\leq i \leq n} V(G_i)$
                and edge set $\bigcup_{1\leq i \leq n}{E(G_i)} \cup \{(v, w) : v \in G_i, w \in G_j, (v_i, v_j) \in E(G)\}$.
                Each graph $G_i$ is substituted for a vertex $v_i$,
                 and all edges between graphs corresponding to adjacent vertices in $G$ are added.
    These operations, taken together in order to construct a graph, 
    form a \emph{parse-tree} of the graph.
    The width of a graph is the maximum size of the vertex set of $G$
     used in operation (O4) to construct the graph.
     The \emph{modular-width} is the minimum width such that
     $G$ can be obtained from some sequence of operations (O1)-(O4).
Finding a parse-tree of a given graph, called a \emph{modular decomposition}, can be done in linear-time \cite{tedder_simple_2007}.

\textit{The second open question is whether we can develop a framework of approximate modular decomposition applicable to real-world datasets?}
Unfortunately, most real-world graphs tend to have larger modular-width.
It would be beneficial if we can efficiently build non-exact parse trees with much lower width but without losing much information.
Possible avenues of exploration include graph editing, a relaxed definition of the parse-tree, and a data-driven approach.

\subsection{Advancing the Parameterized View on Graph Modification}

One of the most explored topics in parameterized complexity are so called \emph{distance to triviality problems} (see, for example,~\cite{DBLP:journals/mst/FominGT20,10.1145/3406325.3451068}). The intuitive question behind these problems is always ``can we make a small change to our input so that it takes on some property?''. In terms of graph problems, one can state a meta-problem as follows, where~$\mathcal{P}$ is a graph property \textit{Vertex-Deletion-To-}$\mathcal{P}$: given a graph $G$, an integer $k$ as well as a parameter $k$, the question is can we delete at most $k$ vertices from $G$, such that the resulting graph has property $\mathcal{P}$?

For many graph properties for which one can consider this meta-problem, either tractable algorithms or complexity lower bounds are known. 
On the other hand, in some application areas it could be useful to delete as many vertices as possible, while ensuring that the resulting graph has a certain property. This leads to the \textit{Max-Vertex-Deletion-To-}$\mathcal{P}$ problem: given a graph $G$, an integer $k$ as well as a parameter $k$, the question is can we delete \emph{at least} $k$ vertices from $G$, such that the resulting graph has property $\mathcal{P}$?

While the change to the problem statement is deceptively simple, we have to this date no complexity-theoretic insight into this class of problems. Note that this problem also differs from the widely used kernelization techniques, as in kernelization, we ask for the resulting input size to be bounded by, for example, $f(k)$. As parameterized complexity can be seen as providing a mathematically rigorous framework of preprocessing through the rich methods of kernelization techniques and algorithmics for distance to triviality problems, extending this framework to further variants of preprocessing seems very natural and could provide further complexity-theoretic and algorithmic insights. These techniques could potentially be useful in the area of (hyper)graph decomposition.

\textbf{Acknowledgments:} We acknowledge Dagstuhl Seminar 23331 on ``Recent Advances in Graph Decomposition''.
\textit{Deepak Ajwani}: Supported in part by a grant from Science Foundation Ireland under Grant number 18/CRT/6183. 
\textit{Johannes Meintrup}: Funded by the Deutsche Forschungsgemeinschaft (DFG, German
Research Foundation) – 379157101. We acknowledge support by DFG grant SCHU 2567/5-1.
\bibliographystyle{abbrv}
\bibliography{references}

\end{document}